\documentclass[twocolumn,showpacs,preprintnumbers,amsmath,amssymb]{revtex4}
\usepackage{amsmath}
\usepackage{graphicx}
\begin{document}

\title{\bf Is the dispersion anomaly of the electron spectrum
induced by the charge-density-wave in high-$T_c$ superconductors?}

\author
{
 T. Zhou$^{1}$ and Z. D. Wang$^{1,2}$
}

\affiliation{$^{1}$Department of Physics and Center of Theoretical
and Computational Physics, The University of Hong Kong, Pokfulam
Road, Hong Kong, China\\
$^{2}$National Laboratory of Solid State Microstructures, Nanjing University, Nanjing 210093, China\\
}

\date{\today}

\begin{abstract}

We propose that the presence of a rotationally symmetric charge
density wave (CDW) with the modulation of $4$-lattice constant in
high-$T_c$ superconductors is essentially responsible for an
anomalous quasiparticle dispersion revealed recently by
angle-resolved photoemission spectroscopy (ARPES) experiments. We
elaborate clearly that the nodal quasiparticle is well-defined at
low energies, while the dispersion breaks up at the energy $E_1$ and
reappears at the energy $E_2$. Our results are in good agreement
with both the ARPES and scanning tunneling microscopy experiments.

\end{abstract}


\maketitle

Recently, scanning tunneling microscopy (STM) experiments have been
performed to search for hidden electronic orders in high-$T_c$
superconductors. A checkerboard pattern with the spatial
four-unit-cell modulation for the local density of states was first
seen around magnetic vortex cores in slightly overdoped samples by
Hoffman $et$ $al$.~\cite{hof}. This kind of modulation pattern was
also reported in later STM experiments, in zero field and in both
the superconducting (SC) and normal states~\cite{vers,how}. In
addition, the Fourier transform of the STM spectra (FT-STM) reveals
four non-dispersive peaks at about $(\pm 2\pi/4,0)$ and $(0,\pm
2\pi/4)$ in the normal~\cite{vers} and SC state~\cite{how}. All of
these experimental results indicate that a kind of charge density
wave (CDW) order with a characteristic wave vector in the CuO bond
direction of $q=2\pi/4$ exists likely in high-$T_c$
superconductors~\cite{zha}.

On the other hand, an intriguing anomalous quasiparticle dispersion
was very recently 
observed in angle-resolved photoemission spectroscopy (ARPES)
experiments~\cite{gra,pan}, i.e., a well-defined nodal quasiparticle
dispersion is broken down between the two higher energies $E_1$ and
$E_2$, which drops in a waterfall-fashion with a very low spectral
weight. Below $E_2$, the quasiparticle dispersion reappears and
disperses towards the zero center. These anomalous features were
observed in several families of cuprates, in under-, optimal- as
well as over-doped samples, and below or above the SC transition
temperature. In hole-doped samples, the energies $E_1$ and $E_2$ are
around $0.35$ and $0.8$ eV, respectively. While in the
electron-doped samples (Pr$_{1-x}$LaCe$_x$CuO$_4$), the
corresponding energies are found to be significantly larger with
$E_1$ being around $0.6$ eV~\cite{pan}. The mechanism of these
features is still unclear while it should be independent of the SC
order. In addition, similar anomalous higher-energy features were
also observed in the insulating cuprate
Ca$_2$CuO$_2$Cl$_2$~\cite{ron}. Thus the mode coupling
picture~\cite{norm} can hardly account for these higher-energy
behaviors because it is expected to happen only in metallic systems.
Note that, it was proposed before that the CDW order  may cause
several other puzzling features seen in ARPES
experiments~\cite{kiv,dama,shen}. Moreover, a connection between the
ARPES and the FT-STM spectra was recently established~\cite{mar,mce}
through the autocorrelation of the ARPES (AC-ARPES) spectra.
Therefore, it is natural and significant to ask whether the
above-mentioned anomalous dispersion observed in ARPES experiments
can also be originated from the CDW order, which has already been
detected by the STM experiments. We here answer this question
clearly by studying the spectral function and the FT-STM based on a
simple phenomenological model including the CDW order, though we
note that a recent theory based on a new representation of the $t-J$
model with two-band fermions was also proposed for the
anomaly~\cite{wang}.

At present, however, our knowledge on the origin and effects of the
CDW order is still far from complete. What the phase diagram of such
an order is and whether it exists in all cuprate materials are still
unclear~\cite{kiv}. In fact, the STM experiments are mostly carried
out on the Bi$_2$Sr$_2$CaCu$_2$O$_{8+x}$ and
Ca$_{2-x}$Na$_x$CuO$_2$Cl$_2$ samples. While a similar
four-unit-cell structure was also found in the YBa$_2$Cu$_3$O$_y$
samples by diffuse x-ray scattering measurements~\cite{zah}. On the
other hand, it seems widely believed that the CDW order may have the
same physics as the stripe order~\cite{kiv}, which has been seen in
the La$_{2-x}$Sr$_x$CuO$_4$ sample~\cite{tra}. In addition, the
stripe order is suggested to exist in various doping region and to
be the ground state of the cuprates~\cite{machi}. Considering the
above experimental and theoretical results, in this paper, we employ
a phenomenological scenario that admits the existence of the CDW
order to look into and elaborate its effect on the spectral
function.  The FT-STM spectra are also evaluated within our simple
model and are compared with the STM experiments. We demonstrate that
the CDW scattering is able to naturally lead to the higher-energy
anomaly in ARPES experiments and the non-dispersive peaks in the
FT-STM spectra. Considering the robustness of this higher-energy
anomaly, our results suggest that the CDW order be likely more
robust in high-$T_c$ superconductors.

We start with the following phenomenological Hamiltonian with a CDW
order being taken into account,

\begin{eqnarray}
H=H_{BCS}+H_{CDW},
\end{eqnarray}
where $H_{BCS}$ is the BCS-type Hamiltonian given by
\begin{eqnarray}
 H_{BCS}=\sum_{\bf k,\sigma}\varepsilon_{\bf
k}c^{\dagger}_{{\bf k}\sigma}c_{{\bf k}\sigma}+\sum_{\bf
k}(\Delta_{\bf k}c^{\dagger}_{\bf k\uparrow}c^{\dagger}_{-{\bf
k}\downarrow}+h.c.),
\end{eqnarray}
 with $\varepsilon_{\bf k}=-2t(\cos k_x+\cos k_y)-4t^{\prime}\cos
k_x\cos k_y-\mu$, $\Delta_{\bf k}=\Delta_0 (\cos k_x-\cos k_y)/2$,
and $H_{CDW}$ is the CDW scattering term,
\begin{equation}
H_{CDW}=\sum_{{\bf Q}{\bf k}\sigma}(Vc^{\dagger}_{{\bf k}+{\bf
Q}\sigma}c_{{\bf k}\sigma}+h.c.),
\end{equation}
with ${\bf Q}=(0,2\pi/4)$ and $(2\pi/4,0)$ being the CDW scattering
wave vectors. Hereafter, the related parameters are chosen as
$t=0.35$ eV, $t^{\prime}=-0.25t$. The chemical potential $\mu$ is
determined by the doping density $\delta=0.12$. We here use the
isotopic $s$-symmetry CDW with the intensity $V=0.1$
eV~\cite{note2}. We have examined that our results are not sensitive
to the slight changes of the chosen parameters and the robustness of
our main conclusion with respect to the CDW scattering intensity
$V$.

\begin{figure}

\centering
\includegraphics[scale=0.46]{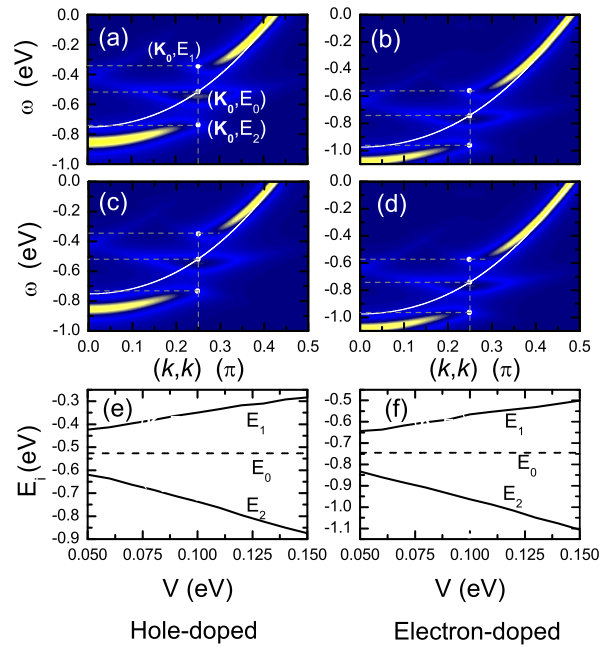}
\caption{(Color online) Panels (a-d) are the intensity plots of the
spectral function as functions of the momentum and energy in the
normal state (a,b) and SC state (c,d), respectively. The solid white
lines represent the bare dispersions $\varepsilon_{\bf k}$. Panels
(e,f) are the energies $E_i$ versus the CDW scattering intensity $V$
in the normal state (see the text). The left and right panels are
for the hole- and electron-doped cases, respectively.
 }\label{fig1}
\end{figure}

The electronic spectral function 
is given by $A({\bf k},\omega)=-1/{\pi}$Im$G({\bf
k},\omega+i\Gamma)$, where the retarded Green's function $G({\bf k},
\omega+i\Gamma)$ is obtained by diagonalizing the Hamiltonian
[Eq.(1)]. The intensity plots of $A({\bf k},\omega)$ are presented
in Figs.1(a-d). We also plot the bare quasiparticle dispersions
$\varepsilon_{\bf k}$ in the absence of the CDW scattering (the
solid white lines) as a function of the momentum ${\bf k}$ for
comparison. As seen, at low energies, the quasiparticle dispersions
 coincide almost with the bare ones. While they deviate from the bare
ones when the momentum is close to ${\bf K_0}=(\pi/4,\pi/4)$. Then
they break up at the energy $E_1$ around the momentum ${\bf K_0}$
and the quasiparticles would be ill-defined within $E_1<E<E_2$. At
the energy $E_2$ the dispersions reappear and disperse towards the
$(0,0)$ point. These results are independent on the SC order. While
the energy $E_i$'s for the electron-doped cuprates are larger than
the corresponding ones for the hole-doped samples, namely,
$E_1\approx0.35$ eV, $E_2\approx0.75$ eV in the hole-doped ones
while $E_1\approx0.6$ eV, $E_2\approx0.95$ eV in the electron-doped
ones. These results are well consistent with the
experiments~\cite{gra,pan}. $E_i$ as a function of the CDW
scattering intensity $V$ is plotted in Figs.1(e) and (f). $E_0$ is
the bare quasiparticle energy at the momentum ${\bf K_0}$. $E_1$ and
$E_2$ approach to $E_0$ as $V$ approaches to zero. As $V$ increases,
$E_1$ decreases and $E_2$ increases linearly. The main results are
stable and robust as $V$ changes slightly.

A sound explanation for the described anomalous higher-energy
features can be given based on the following scattering
characteristic illustration. The first Brillouin zone as well as the
normal state Fermi surfaces of the electron- and hole-doped cuprates
are presented in Fig.2(a). As seen, the Brillouin zone may be
divided into sixteen parts in the presence of the CDW term, labeled
as $'1'$ to $'16'$. When the CDW term acts on an electron in part
$'i'$, it makes the electron hop to the four neighboring
parts~\cite{note}. Let us assume that an electron at $A_0$ in region
$'1'$ has the momentum $(k,k)$ (i.e., along the diagonal direction).
The effect of the CDW term makes it hop to $A_{1-4}$ in the
neighboring regions $'8,4,2,11'$, respectively, as shown in
Fig.2(a). The corresponding momentums for the electrons at $A_{1-4}$
are $(k-\pi/2,k), (k,k+\pi/2), (k+\pi/2,k), (k,k-\pi/2)$,
respectively. We present the bare dispersions for $A_i$ $(i=0-4)$ in
Fig.2(b). The dispersions for the electrons at $A_1$ and $A_4$ (or
$A_2$ and $A_3$) coincide because they are symmetric points. We can
see from Fig.2(b) that the quasiparticle energy at $A_0$ is
different from that at $A_i$ as the momentum is far from ${\bf
K_0}$. Thus the hopping from $A_0$ to $A_i$ is difficult to occur
because the energy conservation condition is not satisfied. As a
result, the CDW term has little effect on the hopping for this case,
namely, the quasiparticle is well-defined and the dispersion
coincides almost with the bare one, as shown in Fig.1(a). When the
momentum decreases and moves close to ${\bf K_0}$, the energy at
$A_0$ is closer to that at $A_1$ and $A_4$. The CDW term influences
significantly the dispersion. Thus it deviates from the bare one and
the spectral weight reduces gradually. When the momentum reaches
${\bf K_0}$, the bare quasiparticle energy at $A_0$ just equals to
that at $A_1$ and $A_4$, as seen in Fig.2(b), so that the hopping
from $A_0$ to $A_{1,4}$ can occur easily, which destroys the state
of the quasiparticle at $A_0$ and leads the dispersion to break up
at this momentum. As a result, the dispersion is pinned by the CDW
term and the quasiparticle is ill-defined. While,  the quasiparticle
dispersion is no longer pinned by the CDW term when the bounding
energy is high enough (larger than $E_2$), and thus it reappears and
disperses towards $(0,0)$ point. In fact, when the electron leaves
the ${\bf K_0}$ point, the energy at $A_0$ is different from that at
$A_i$ and thus the quasiparticle is well-defined again.

\begin{figure}

\centering

\includegraphics[scale=0.5]{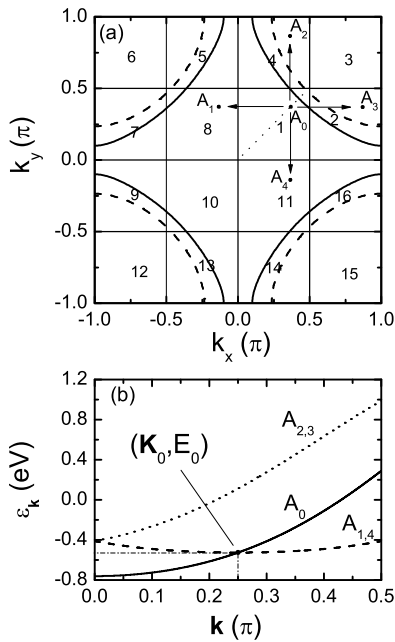}
\caption{(a) The normal state fermi surfaces for the hole-doped
(solid line) and electron-doped (dashed line) cuprates,
respectively. The scattering from $A_0\rightarrow A_i$ indicates the
hopping caused by the CDW term. (b) The bare dispersion of
quasiparticles at $A_0$ (along the diagonal direction) and the
corresponding dispersions of that at $A_i$ for the hole-doped case.
}\label{fig2}
\end{figure}

In the SC state, the quasiparticle energy at $A_0$ is the same as
that in the normal state along the diagonal direction, while that at
$A_i$ is changed due to the presence of the SC gap. Meanwhile, the
quasiparticle energy at $A_0$ equals to that at $A_{1,4}$ at the
momentum ${\bf K_0}$ because $A_{1,4}$ is along the nodal direction
at this momentum. Thus the addressed anomaly depends weakly on the
SC order parameter $\Delta_0$. For the electron-doped cases, since
the Fermi momentum ${\bf K_F}$ along the diagonal direction is
greater than that of the holed-doped sample, as seen from the Fermi
surfaces shown in Fig.2(a), the bounding energy at ${\bf K_0}$ is
larger. As a result, the anomalous energies $E_1$ and $E_2$ are
larger than those of the hole-doped ones, consistent with the
experimental results~\cite{pan}.

\begin{figure}

\centering

\includegraphics[scale=0.6]{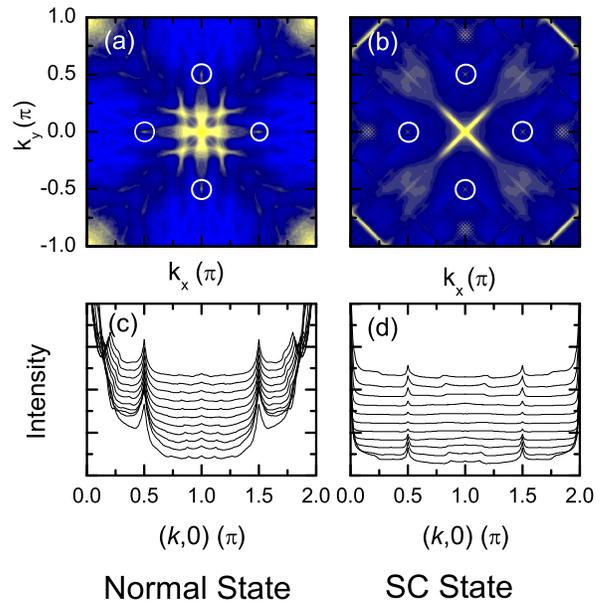}
\caption{(Color online) Panels (a) and (b) are the intensity plots
of the AC-ARPES spectra $C({\bf q},\omega)$ at the zero energy in
the normal and SC state ($\Delta_0=0.04$ eV), respectively. White
circles are used to denote the energy-independent peaks. Panels (c)
and (d) are the two-dimensional cuts for the spectra along $(0,0)$
to $(2\pi,0)$ with the energy increasing from $-0.02$ eV to $0.02$
eV (from the bottom to the top) in the normal and SC state,
respectively. }\label{fig3}
\end{figure}

We now turn to address the FT-STM spectra in hole-doped cuprates. As
we mentioned above, the FT-STM spectra can be related to the ARPES
spectra through the AC-ARPES function $C({\bf q},\omega)$, given
by~\cite{mar},
\begin{equation}
C({\bf q},\omega)=\frac{1}{N}\sum_{\bf k}A({\bf k},\omega)A({\bf
k}+{\bf q},\omega).
\end{equation}
Very recently, it was concluded experimentally that the FT-STM
spectra are in good agreement with the AC-ARPES spectra at low
energies for various doping densities~\cite{mce}. Therefore, we here
use the AC-ARPES to deduce the FT-STM spectra and compare the
results with the STM experiments. The calculated intensities of the
AC-ARPES spectra at zero energy, in the normal and SC state, are
shown in Fig.3(a) and Fig.3(b), respectively. The log scale is used
so that the weak features can be revealed. As shown, four peaks
appear at the momentums $(\pm2\pi/4,0)$ and $(0,\pm2\pi/4)$. These
peaks are independent of the energies, as seen more clearly in
Figs.3(c) and (d), which are the two-dimensional cuts of the spectra
for the energies from $-0.02$ to $0.02$ eV. These non-dispersive
peaks in the FT-STM spectra were observed experimentally in the
normal state~\cite{vers}. In the SC state, while they are suppressed
by the SC order, they were also indeed observed by STM experiments
as well~\cite{how}.

\begin{figure}

\centering
\includegraphics[scale=0.45]{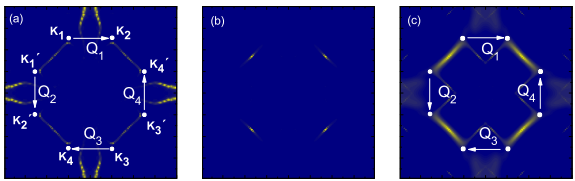}
\caption{(Color online) Panels (a) and (b) are the intensity plots
of ARPES spectra at zero energy in the normal and SC state,
respectively. Panel (c) is the replotting of the spectra in the SC
state while the log scale is used.}\label{fig4}
\end{figure}

The consistency between the FT-STM and AC-ARPES spectra indicates
that the non-dispersive peaks may be explained based on the
quasiparticle interference model~\cite{qwang,mcel}, i.e., the
momentums of the peaks in FT-STM spectra are just the wave vectors
that connect the tips of the constant energy contour. Based on this
picture, the various dispersive peaks observed by STM experiments in
the SC state~\cite{mcel,hoff} are reproduced successfully. At
present, in the presence of the CDW order, the intensity plots of
the ARPES spectra at zero energy are depicted in Figs.4(a) and
(b)/(c) for the normal state and SC state, respectively; since the
ARPES spectrum is reduced to a dot along the nodal direction in the
SC state [Fig.4(b)], we re-plot it by using a log scale in Fig.4(c)
to reveal the weak features caused by the CDW order. As we discussed
above, the ARPES spectra break up at the momentums ${\bf
K_i}=(\pm\pi/4,q)$ and ${\bf K^{\prime}_i}=(q,\pm\pi/4)$ because of
the CDW scattering, leading to the tips of the spectra at these
momentums. In the corresponding AC-ARPES spectra, the peaks appear
at the momentums ${\bf Q_i}$ that connect the tips of the ARPES
spectra, as denoted in Figs.4(a) and (c). As a result, although the
CDW order has a very weak effect on the low energy ARPES spectra,
which is difficult to be detected directly by the ARPES experiments,
the non-dispersive peaks induced by the CDW order appear clearly in
the AC-ARPES spectra (or the FT-STM spectra) and can be detected
firmly by the STM experiments.

To conclude, based on a phenomenological model, we have for the
first time elaborated that the CDW scattering is able to cause the
intriguing dispersion anomaly observed in the ARPES spectra for
both hole-doped and electron-doped samples as well as the
non-dispersive peaks in the FT-STM spectra. A clear physical
picture of the mechanism was presented based on the scattering
analysis in the presence of the CDW order.

We thank Y. Chen, L. H. Tang, and Q. H. Wang for helpful
discussions. This work was supported by the RGC grants of Hong
Kong (HKU 7050/03P and HKU-3/05C), the NSFC (10429401), and the
973 project of China (2006CB601002).


\begin{thebibliography}{99}
\bibitem{note2} A further study for the symmetry of the CDW
may agree better with experiments,  while it is not a main concern
of the present paper.
\bibitem{note} Taking into account the periodicity in the momentum space,
the parts that including the border of the Brillouin zone, such as
parts $2$ and $7$, are also seen as the neighbor parts.
\end{thebibliography}
\end{document}